\documentclass[prd]{revtex4}
\usepackage{graphicx}
\usepackage{dcolumn}
\usepackage{bm,color}
\usepackage{hyperref}

\usepackage{amssymb,float}
\usepackage{amsmath}
\usepackage[utf8]{inputenc}
\usepackage[dvipsnames]{xcolor}
\usepackage{enumerate}

\newcommand{\udt}[3]{#1^{#2}_{\phantom{#2}#3}}

\newcommand{\dut}[3]{#1_{#2}^{\phantom{#2}#3}}
\newcommand{\dudt}[4]{#1_{#2\phantom{#3}#4}^{\phantom{#2}#3}}
\begin{document}

\title{Reviving Horndeski Theory using Teleparallel Gravity after GW170817}
\author{Sebastian Bahamonde}
\email{sbahamonde@ut.ee, sebastian.beltran.14@ucl.ac.uk}
\affiliation{Laboratory of Theoretical Physics, Institute of Physics, University of Tartu, W. Ostwaldi 1, 50411 Tartu, Estonia}
\affiliation{Department of Mathematics, University College London, Gower Street, London, WC1E 6BT, United Kingdom}

\author{Konstantinos F.	Dialektopoulos}
\email{dialektopoulos@na.infn.it}
\affiliation{Center for Gravitation and Cosmology, College of Physical Science and Technology, Yangzhou University, Yangzhou 225009, China,}
\affiliation{Aristotle University of Thessaloniki, Thessaloniki 54124, Greece}
\affiliation{Institute of Space Sciences and Astronomy, University of Malta, Msida, Malta}

\author{Viktor Gakis}
\email{vgakis@central.ntua.gr}
\affiliation{Institute of Space Sciences and Astronomy, University of Malta, Msida, Malta}
\affiliation{Department of Physics, National Technical University of Athens, Zografou Campus GR 157 73, Athens, Greece}

\author{Jackson Levi Said}
\email{jackson.said@um.edu.mt}
\affiliation{Institute of Space Sciences and Astronomy, University of Malta, Msida, Malta}
\affiliation{Department of Physics, University of Malta, Msida, Malta}

\date{\today}

\begin{abstract}
Horndeski gravity was highly constrained from the recent gravitational wave observations by the LIGO Collaboration down to $|c_{g}/c-1|\gtrsim 10^{-15}$. In this paper, we study the propagation of gravitational waves in a recently proposed model of Horndeski gravity in which its Teleparallel Gravity analog is formulated. As usually done in these analyses, we consider a flat cosmological background in which curvature is replaced by torsion as the expression of gravitation. It is found that in this approach, one can construct a more general Horndeski theory satisfying $c_T=c_g/c=1$ without eliminating the coupling functions $G_5(\phi,X)$ and $G_4(\phi,X)$ that were highly constrained in standard Horndeski theory. Hence, in the teleparallel approach one is able to restore these terms, creating an interesting way to revive Horndeski gravity. In this way, we retain the original spirit of Horndeski gravity (unlike beyond Horndeski theories) while only changing the form in which the geometry of gravitation is expressed.
\end{abstract}

\maketitle



\section{Introduction}
The binary neutron star merger events associated with the gravitational wave (GW) GW170817 \cite{TheLIGOScientific:2017qsa} and its companion electromagnetic counterpart GRB170817A \cite{Goldstein:2017mmi} has tremendously constrained the GW speed of propagation to the speed of light to within deviations of at most one part in $10^{15}$. The birth of multimessenger GW astronomy has thus placed a dramatic constraint on models of gravity predicting deviations in this difference of propagation speeds. One such theory is Horndeski gravity \cite{Horndeski:1974wa} which is the most general second-order theory of gravity involving a single scalar field in four dimensions. Horndeski gravity has been used in a diverse range of settings but is particularly useful for constructing models of inflation and dark energy \cite{Kobayashi:2011nu,Gleyzes:2013ooa,Koyama:2015vza} (and references therein).

Horndeski was able to write his theory of gravity in closed form because of the appearance of Lovelock's theorem~\cite{Lovelock:1971yv} which states that in four dimensions, the only possible second-order theory of gravity is general relativity (GR), up to an integration constant (satisfying also reasonable conditions such as diffeomorphism and Lorentz invariance). Together with the finite contribution of the scalar field, Horndeski gravity provides a concise general framework on which to construct second-order theories of gravity. However, the speed of propagation of GWs in Horndeski gravity~\cite{Nunes:2018zot} has severely limited the potential models of the theory~\cite{Creminelli:2017sry,Sakstein:2017xjx}. While the format of the theory has not been narrowed to GR, its most cosmologically interesting models have been eliminated or severely limited. Moreover, it is important to explore possible ways to revive Horndeski gravity because the majority of modified gravity theories feature as subclasses of the fuller Horndeski theory~\cite{Bellini:2014fua}. This has prompted a resurgence in work refining the central theme of Horndeski gravity and has then led to beyond Horndeski gravity models~\cite{Kobayashi:2019hrl}, in which the second-order field equation condition is relaxed but where the Ostrogradski ghosts are removed. Another possible avenue to revive Horndeski gravity is in the context of effective field theories where Horndeski gravity may exist as the classical limit within some larger UV complete theory. This may allow for a frequency-dependent speed of propagation for GWs such as in Ref.\cite{deRham:2018red}.

Horndeski gravity offers a platform on which to construct modified theories of gravity in which several general results can be safely inherited. These range from the fact that all subclasses of Horndeski gravity are ghost free and observe the weak equivalence principle \cite{Kobayashi:2011nu} to them having only one extra propagating scalar degree of freedom in addition to those that appear in GR. Also, Horndeski gravity encapsulates the common features and behaviors that adding one scalar field to a theory and demanding that the ensuing field equations remain second-order in nature result. An example of this comes from large-scale structure in the Universe where the linear growth rate of structure is systematically lower than its $\Lambda$CDM counterpart while this is larger for higher redshifts \cite{Perenon:2015sla}. Likewise, Horndeski gravity offers a natural relationship between an early-time inflationary epoch \cite{Kobayashi:2019hrl} and a late-time dark energy behavior \cite{Clifton:2011jh}. The effect brought to the fore by Horndeski gravity is also impactful at the galactic scale where the scalar field acts dynamically by mimicking dark matter \cite{Rinaldi:2016oqp,Chan_2018,SpurioMancini:2019rxy}. For these reasons and others, Horndeski gravity can offer a reasonable area from which to produce new cosmological models. Our work is focused on reviving this approach to cosmology while respecting recent multimessenger observations on the GW speed of propagation.

Horndeski's theory of gravity assumes outright that gravity is described by the Levi-Civita connection which is the basis of GR and the vast majority of modified gravity~\cite{Clifton:2011jh}. The Levi-Civita connection is torsionless, satisfies the metric compatibility condition, and describes gravitation by means of a curvatureful Riemann tensor~\cite{misner1973gravitation}. On the other hand, Teleparallel Gravity (TG) formulated on the Weitzenb\"{o}ck connection is curvatureless and torsionful~\cite{Aldrovandi:2013wha} (and continues to satisfy the metric compatibility condition). One benefit of TG is that its analog of Lovelock's theorem is not bounded in terms of Lagrangian contributions~\cite{Gonzalez:2015sha}. This means that Lovelock's theorem alone will produce an infinite number of terms in the TG Lagrangian. The consequence of this property is that the TG analog of Horndeski gravity grants another route to producing an observationally consistent theory that retains the spirit of Horndeski gravity.

In Ref.~\cite{PhysRevD.100.064018}, Bahamonde--Dialektopoulos--Levi Said (BDLS) developed the details of this theory under reasonably physical conditions (that will be explained later on). The product is a new Lagrangian component in addition to those that appear in the original version of Horndeski gravity. BDLS theory opens a new possibility to revive Horndeski gravity within the TG context. This will raise previously eliminated models to subclasses of the newly proposed theory where the new TG component will be constrained through observational tests.

In this paper, we first review the newly proposed BDLS theory in Sec.~\ref{BDLS_Rev}, and then show that the propagation of tensor modes in BDLS theory can resurrect many of the disqualified models of standard Horndeski gravity in Sec.~\ref{GWPE_Derivation}. This is done by determining the speed of propagation of gravitational waves. In Sec.~\ref{Reviving_HG_Examples}, we present examples of how this can be done for some interesting models that are inspired by standard Horndeski theory. Finally, we close in Sec.~\ref{Conclusion} with a summary and conclusions of the core results. Throughout this work, we take units in which the speed of light is equal to unity unless otherwise stated.

\section{The Teleparallel Gravity Analog of Horndeski Gravity (BDLS theory)\label{BDLS_Rev}}
GR expresses gravitation by means of the metric tensor $g_{\mu\nu}$ through the Levi-Civita connection $\mathring{\Gamma}^{\sigma}_{\mu\nu}$ (we use overcircles throughout to denote quantities determined by the Levi-Civita connection). This is measured via the Riemann tensor which gives a meaningful measure of curvature in standard gravity (meaning theories based on the Levi-Civita connection), and is used in the construction of many extended theories of gravity. On the other hand, the fundamental dynamical object of TG is the tetrad $\udt{e}{a}{\mu}$ which acts as a soldering agent between the local Minkowski space (Latin indices) and the general manifold (Greek indices).

The tetrads reproduce the metric through
\begin{equation}\label{metric_def}
    g_{\mu\nu} = \udt{e}{a}{\mu}\udt{e}{b}{\nu}\eta_{ab}\,,
\end{equation}
and observe the inverse transformation relation 
\begin{equation}
    \eta_{ab} = \dut{e}{a}{\mu}\dut{e}{b}{\nu}g_{\mu\nu}\,.
\end{equation}
Also, the tetrad is normalized by the orthogonality relations
\begin{align}
    \udt{e}{a}{\mu}\dut{e}{b}{\mu} &= \delta^a_b\,,\\ \udt{e}{a}{\mu}\dut{e}{a}{\nu} &= \delta^{\nu}_{\mu}\,.
\end{align}
Consequently, there is an infinite set of tetrads that satisfy these conditions. TG theories are based upon the Weitzenb\"{o}ck connection, which is curvatureless and metric compatible. The linear affine form of this connection can be related to its spin connection counterpart through the relation
\begin{equation}
    {\Gamma}^{\sigma}{}_{\mu\nu} = \dut{e}{a}{\sigma}\partial_\mu \udt{e}{a}{\nu} + \dut{e}{a}{\sigma}\udt{\omega}{a}{b\mu}\udt{e}{b}{\nu}\,.
\end{equation}
As in GR, the spin connection $\udt{\omega}{a}{b\mu}$ accounts for the local Lorentz transformation (LLT) degrees of freedom, but in TG this plays an active role in the equations of motion of the theory by offsetting inertial effects that arise from the freedom in choosing the tetrad, i.e. solutions to Eq.(\ref{metric_def}). In any setting, one can always choose the so-called purely inertial gauge in which the spin connection vanishes organically due to an appropriate choice of frame~\cite{Krssak:2018ywd}. The purely inertial gauge can also be seen as the Lorentz frame in which the spin connection components vanish.

By choosing the Weitzenb\"{o}ck connection, the Riemann tensor identically vanishes, whereas the torsion tensor defined by~\cite{Aldrovandi:2013wha}
\begin{equation}
\udt{T}{a}{\mu\nu} :=  2\Gamma^{a}_{[\mu\nu]}\,,
\end{equation}
quantifies the field strength of gravity in TG. This quantity can be decomposed into irreducible axial, vector and purely tensorial parts defined, respectively, as~\cite{Bahamonde:2017wwk}
\begin{align}
a_{\mu} &= \frac{1}{6}\epsilon_{\mu\nu\sigma\rho}T^{\nu\sigma\rho}\,, \\
v_{\mu} &= \udt{T}{\sigma}{\sigma\mu}\,,\\
t_{\sigma\mu\nu} &= \frac{1}{2}\left(T_{\sigma\mu\nu}+T_{\mu\sigma\nu}\right) + \frac{1}{6}\left(g_{\nu\sigma}v_{\mu} + g_{\nu\mu}v_{\sigma}\right) - \frac{1}{3}g_{\sigma\mu}v_{\nu}\,,
\end{align}
where $\epsilon_{\mu\nu\sigma\rho}$ is the totally antisymmetric Levi-Civita symbol in four dimensions. These are irreducible parts with respect to the local Lorentz group and can be used to construct scalar invariants
\begin{align}
T_{\text{ax}} &= a_{\mu}a^{\mu} = \frac{1}{18}\left(T_{\sigma\mu\nu}T^{\sigma\mu\nu} - 2T_{\sigma\mu\nu}T^{\mu\sigma\nu}\right)\,,\\
T_{\text{vec}} &=v_\mu v^\mu= \udt{T}{\sigma}{\sigma\mu}\dut{T}{\rho}{\rho\mu}\,, \\
T_{\text{ten}} &= t_{\sigma\mu\nu}t^{\sigma\mu\nu} = \frac{1}{2}\left(T_{\sigma\mu\nu}T^{\sigma\mu\nu} + T_{\sigma\mu\nu}T^{\mu\sigma\nu}\right)- \frac{1}{2}\udt{T}{\sigma}{\sigma\mu}\dut{T}{\rho}{\rho\mu}\,.
\end{align}
These three quantities form the most general second-order Lagrangian density that is quadratic in the torsion tensor and is parity preserving \cite{PhysRevD.19.3524}, which can be written as $f(T_{\text{ax}},T_{\text{vec}},T_{\text{ten}})$. For the special choice of linear coefficients
\begin{equation}
T = \frac{3}{2} T_{\text{ax}} + \frac{2}{3} T_{\text{ten}} - \frac{2}{3} T_{\text{vec}}\,,
\end{equation}
the resulting Lagrangian turns out to be equivalent to the Ricci scalar $\mathring{R}$ (computed with the Levi-Civita connection) up to a total divergence term~\cite{Bahamonde:2015zma}
\begin{equation}\label{Ricc_Tor_iden}
    \mathring{R}=-T+B\,,
\end{equation}
where $B$ is a boundary contribution. This is the so-called \textit{teleparallel equivalent of general relativity} and results in identical field equations as GR, despite differing at the level of the action.

The procedure to transform local Lorentz frames to the general manifold in GR comprises of exchanging the Minkowski metric for the general manifold metric tensor and raising the partial derivative to the Levi-Civita covariant derivative. In TG, the Minkowski manifold is formed by trivial tetrads. The coupling procedure for a general scalar field, $\Psi=\Psi(x^a(x^{\mu}))$, is then prescribed by elevating these trivial tetrads to general tetrads, $\udt{e}{a}{\mu}$, and by mapping the derivative operator through~\cite{Krssak:2018ywd}
\begin{equation}
\partial_{\mu}\,\rightarrow\,\mathring{\nabla}_{\mu}\,,
\end{equation}
where the action of this operator retains the same form as in GR which is a result of the close relationship the two theories share.

Now that both the gravitational and scalar field sectors have been adequately developed, we can lay the criteria on which to construct the TG analog of Horndeski gravity in four dimensions~\cite{PhysRevD.100.064018}, which are 
\begin{enumerate}[(i)]
    \item the resulting field equations must, at most, be second order in terms of tetrad derivatives;
    \item the scalar invariants cannot be parity violating; and
    \item contractions of the torsion tensor can be at most quadratic.
\end{enumerate}
The last condition acts to limit the potentially infinite higher-order contractions that may appear in the theory. This is a result of the weakened Lovelock theorem in TG \cite{Gonzalez:2015sha,PhysRevD.100.064018} which now produces a potential infinite number of terms.

Observing these conditions leads directly to the scalar invariants which are linear in the torsion tensor
\begin{equation}
    I_2=v^{\mu}\phi_{;\mu},
\end{equation}
and quadratic in the torsion tensor
\begin{align}
    J_1&=a^{\mu}a^{\nu}\phi_{;\mu}\phi_{;\nu}\,,\\ J_3&=v_{\sigma}t^{\sigma\mu\nu}\phi_{;\mu}\phi_{;\nu}\,,\\ J_5&=t^{\sigma\mu\nu}\dudt{t}{\sigma}{\bar{\mu}}{\nu}\phi_{;\mu}\phi_{;\bar{\mu}}\,,\\ J_6&=t^{\sigma\mu\nu}\dut{t}{\sigma}{\bar{\mu}\bar{\nu}}\phi_{;\mu}\phi_{;\nu}\phi_{;\bar{\mu}}\phi_{;\bar{\nu}}\,,\\
    J_8&=t^{\sigma\mu\nu}\dut{t}{\sigma\mu}{\bar{\nu}}\phi_{;\nu}\phi_{;\bar{\nu}}\,,\\ J_{10}&=\udt{\epsilon}{\mu}{\nu\sigma\rho}a^{\nu} t^{\alpha\rho\sigma}\phi_{;\mu}\phi_{;\alpha}\,,
\end{align}
where the semicolon represents the Levi-Civita covariant derivative. While other permutations exist, they can be shown to reduce to these terms when the symmetries of the torsion tensor are taken into account.

Defining the kinetic term of the scalar field as $X := -\frac{1}{2}\partial^{\mu}\phi\partial_{\mu}\phi$ results in the new Lagrangian component
\begin{align}\label{Ltele}
    \mathcal{L}&_{\text{Tele}} := G_{\text{Tele}}(\phi,X,T,T_{\text{ax}},T_{\text{vec}},I_2,J_1,J_3,J_5,J_6,J_8,J_{10})\,.
\end{align}
By virtue of the TG coupling prescription, the Lagrangian components of Horndeski's theory in standard gravity remain identical except that they are expressed in terms of the tetrad. This means that the TG analog of Horndeski's theory can be written as~\cite{PhysRevD.100.064018}
\begin{equation}\label{BDLS_Action}
    \mathcal{S}_{\text{BDLS}} = \frac{1}{2\kappa^2}\int  d^4x\,  e \mathcal{L}_{\rm Tele}  +\frac{1}{2\kappa^2} \sum_{i=2}^{5}\int d^4x\, e\mathcal{L}_{i}\,,
\end{equation}
where 
\begin{align}
\mathcal{L}_2 &:= G_2(\phi,X)\,,\quad\mathcal{L}_3 := G_3(\phi,X)\Box\phi\,,\label{HG_23}\\
 \mathcal{L}_4 &:= G_4(\phi,X)\left(-T+B\right) + G_{4,X}(\phi,X)\left[\left(\Box\phi\right)^2 - \phi_{;\mu\nu}\phi^{;\mu\nu}\right]\,,\label{HG_4}\\
 \mathcal{L}_5 &:= G_5(\phi,X)\mathcal{G}_{\mu\nu}\phi^{;\mu\nu} - \frac{1}{6}G_{5,X}(\phi,X)\Big[\left(\Box\phi\right)^3 + 2\dut{\phi}{;\mu}{\nu}\dut{\phi}{;\nu}{\alpha}\dut{\phi}{;\alpha}{\mu} - 3\phi_{;\mu\nu}\phi^{;\mu\nu}\left(\Box\phi\right)\Big]\,,\label{HG_5}
\end{align}
in which $\kappa^2=8\pi G$, $e:=\det(\udt{e}{a}{\mu})=\sqrt{-g}$ is the determinant of the tetrad, $\mathcal{G}_{\mu\nu}$ is the regular Einstein tensor, commas denote differentiation and $\Box\phi := \udt{\phi}{;\mu}{;\mu}$. Clearly, for the choice of $G_{\text{Tele}}=0$, standard Horndeski gravity is recovered without exception. Due to the local Lorentz invariance of the torsion tensor, the new BDLS formulation of Horndeski gravity is covariant under both Lorentz transformations and diffeomorphisms.

Given the lower-order nature of TG compounded by the weakened realization of Lovelock's theorem, the TG analog of Horndeski's original theory has a much larger parameter space in TG \cite{PhysRevD.100.064018}. As expected, we recover the standard gravity Horndeski terms in Eqs.(\ref{HG_23})--(\ref{HG_5}) which are now complemented by the additional Lagrangian contribution of $\mathcal{L}_{\rm Tele}$ which appears naturally as part of the TG analog of Horndeski theory \cite{Horndeski:1974wa}. In this new TG analog formulation of Horndeski gravity, we demonstrate one possible approach to reviving Horndeski gravity without resorting to beyond Horndeski theories or other considerations.

In the Appendix we present the field equations of the action in Eq.(\ref{BDLS_Action}) by taking variations with respect to the tetrad and scalar field. In GR, a variation with respect to the metric tensor produces ten independent field equations (after symmetries are taken into account). Similarly, TG produces ten independent field equations from this variation. However, due to the symmetry of the energy-momentum tensor, it was noticed in Ref.\cite{Li:2010cg} that an extra six independent field equations are produced. These represent the field equations due to the invariance of the theory under LLTs and correlate to the six Lorentz transformations. The extra six field equations are produced by considering the antisymmetric operator on the two free indices of the field equations (when considered transformed to their general manifold expression). Since the energy-momentum tensor is symmetric, these equations must vanish for a consistent inertial structure of the dynamical equations. By choosing appropriate the spin connection components \cite{Aldrovandi:2013wha,Krssak:2015oua}, these equations can always be satisfied in the purely inertial gauge these components vanish due to the choice in the particular form of the tetrad components.

\section{The GW Propagation Equation \label{GWPE_Derivation}} 

One of the defining features of GR is that by taking tensor perturbations, it leads directly to a wave equation which can be related to how GWs propagate. In turn, this GW propagation equation (GWPE) can be used to relate various modifications to the gravity section against each other \cite{PhysRevLett.122.061301}, as well as against GR. Recent observations limit the speed of propagation of GWs to that of light to within a highly constrained margin \cite{TheLIGOScientific:2017qsa}
\begin{equation}
\Big|\frac{c_{g}}{c}-1\Big|\gtrsim10^{-15}\,.\label{GW_constraint}
\end{equation}
In fact, in Ref.\cite{Abbott:2016blz} this was used to set an upper bound of the graviton mass to $m_{g}<1.2\times10^{-22}$ eV/$c^{2}$, which leaves little room for a massive graviton. A much stronger constraint exists from Solar System tests in which $m_{g}<10^{-30}$ eV/$c^{2}$ which means that GW170817 did not significantly alter our picture of the graviton \cite{deRham:2016nuf}. For these reasons, any modified theory of gravity must produce GWs that propagate very close to $c$ and predict, at most, a minuscule mass for the associated graviton mass.

In GR, the GWPE emerges through taking tensor perturbations about a background cosmology through $g_{\mu\nu} \rightarrow g_{\mu\nu} + \delta g_{\mu\nu}$, where $|\delta g_{\mu\nu}|\ll 1$ and $g_{\mu\nu}$ represents the background cosmology. In this case, $\delta g_{\mu\nu}$ will carry the GW degrees of freedom (DOF) which in GR is exhibited as two DOFs as part of a massless spin-2 field.

The perturbative approach that appears in theories which are wholly based on the metric tensor $g_{\mu\nu}$ is easily transferable to tetrad-based theories of gravity. Naturally, we take a perturbation of the background tetrad $\udt{e}{a}{\mu}$ where
\begin{equation}\label{tet_pert}
    \udt{e}{a}{\mu} \rightarrow \udt{e}{a}{\mu} + \delta \udt{e}{a}{\mu}\,,
\end{equation}
such that $|\delta \udt{e}{a}{\mu}|\ll 1$ represents the first-order perturbation of the tetrad. As is well known \cite{peacock1999cosmological}, not all DOFs are independent and by taking gauge choices, these superfluous DOFs can be extirpated in cosmological perturbation analyses. Again, this easily follows for tetrad formulated theories in which the regular gauge choices can be readily adapted for this setting. To see this, consider a spatially flat cosmology $ds^{2}=-dt^{2}+a(t)^{2}(dx^{2}+dy^{2}+dz^{2})$ \cite{misner1973gravitation}, which can be straightforwardly produced by the tetrad choice $\udt{e}{a}{\mu}=\text{diag}(1,a(t),a(t),a(t))$. This is a perfect example of a situation where the inertial spin connection components all vanish meaning that the tetrad is the only actor in the ensuing analysis \cite{Hohmann:2019nat}.

At the first-order perturbative level, the metric tensor perturbations take on the form $g_{\mu\nu}=a^2\delta^i_{\mu}\delta^j_{\nu}h_{ij}$ for spatial $i,j$. These tensor perturbations are transverse, traceless and symmetric. Similar to the background scenario, we can choose tetrad components that produce the identical metric entries while also having a vanishing associated spin connection \cite{Tamanini:2012hg,Li:2010cg}. This is achieved for the choice 
\begin{equation}\label{tet_ten_pert}
\delta e^{k}{}_{\mu}=\frac{1}{2}a\,\delta_{\mu}^{i}\delta^{kj}h_{ij}\,,
\end{equation}
where $i,j,k$ are all spatial. It is through these tensor modes that the GWPE for BDLS theory can be determined.

The most general parametrization of the GWPE on a flat cosmological background in modified gravity takes the form \cite{Saltas:2014dha,Riazuelo:2000fc}
\begin{equation}\label{BDLS_GWPE}
\ddot{h}_{ij}+\left(3+\alpha_{M}\right)H\dot{h}_{ij}-\left(1+\alpha_{T}\right)\frac{k^{2}}{a^{2}}h_{ij}=0\,,
\end{equation}
where dots denote differentiation with respect to cosmic time, $H=\dot{a}/a$ is the Hubble parameter, $\alpha_{M}=\frac{1}{HM_{\ast}^{2}}\frac{dM_{\ast}^{2}}{dt}$ is Planck mass running rate, and $\alpha_{T}=c_{T}^{2} - 1$ is the tensor excess speed. The GWPE in Eq.~(\ref{BDLS_GWPE}) is being considered in its Fourier domain, along with a source-free scenario \cite{Gleyzes:2014dya,Charmousis:2011bf}.

The joint observations of GW170817 and its associated electromagnetic counterpart GRB170817A put stringent constraints on the upper bound of $\alpha_T$ which are not compatible with most major manifestations of standard Horndeski theory. In most popular cosmologically interesting versions of this Horndeski theory, we find a propagation speed that varies significantly from the speed of light \cite{PhysRevLett.122.061301}. While several well-motivated theories exist that consider beyond Horndeski gravity (see \cite{Kobayashi:2019hrl}), the TG analog of Horndeski theory offers an avenue that keeps to the original spirit of the approach \cite{Horndeski:1974wa}.

The values of the parameterization variables that appear in Eq.~\eqref{BDLS_GWPE} depend on the particular theory being investigated. In order to do this for BDLS theory, we must consider the tetrad perturbation laid out through Eq.~(\ref{tet_pert}) where the flat cosmological background is perturbed by the tensor modes that appear in Eq.~(\ref{tet_ten_pert}).  These are then substituted into the field equations and the resulting form is formulated to be comparable to Eq.~(\ref{BDLS_GWPE}). As field equations we use the Euler-Lagrange equations in the minisuperspace for the background variables (scale factor, lapse function and scalar field), and the tensorial perturbation $h_{ij}$. We also consider LLT invariance by taking transformations of our initial tetrad ansatz and confirm that the antisymmetric field equations vanish, which verifies that our tetrad is compatible with a vanishing spin connection. While cumbersome, this procedure can be used to probe the nature of GW propagation in any modified theory of gravity (further details in Refs.~\cite{Bardeen:1980kt,Malik:2008im,Mukhanov:1990me} and references therein). By taking this tensor perturbation prescribed, we find values for the two parameters of the GWPE where the excess tensor speed is given by
\begin{widetext} 
\begin{equation}\label{alpha_T}
\alpha_{T} = \frac{2X}{M_{\ast}^{2}}\left(2G_{4,X}-2G_{5,\phi}-G_{5,X}(\ddot{\phi}-\dot{\phi}H)-2G_{{\rm Tele,J_{8}}}-\frac{1}{2}G_{{\rm Tele,J_{5}}}\right)\,,
\end{equation}
\end{widetext}
and the effective Planck mass is given by 
\begin{align}\label{eff_Planck_mass}
M_{\ast}^{2} &= 2\Big(G_{4}-2XG_{4,X}+XG_{5,\phi}-\dot{\phi}XHG_{5,X}+2XG_{{\rm Tele,J_{8}}}+\frac{1}{2}XG_{{\rm Tele,J_{5}}}-G_{{\rm Tele,T}}\Big)\,,
\end{align}
where comas represent derivatives, and the only nonvanishing contributing scalars to the $G_{\text{Tele}}$ term are $T = 6H^{2}/N^{2}$, $T_{\text{vec}} = -9H^{2}/N^{2}$, and $I_{2} = 3H\dot{\phi}/N^{2}$, while the other scalars all vanish up to perturbative order.

The appearance of the $G_{\text{Tele}}$ term in Eq.~(\ref{BDLS_GWPE}) directly leads to a potentially revised speed of GWs as compared to the standard Horndeski theory. This means that we may revive interesting cosmological models from standard gravity by way of solving for the scenario where GWs propagate at the speed of light, i.e. $\alpha_T=0$. The result is that each standard Horndeski model from standard gravity now reemerges as a family of solutions of this new constraint. Naturally, for the situation where $G_{\text{Tele}}=0$ we recover the standard gravity results for the GWPE \cite{Kobayashi:2011nu}. The action of Eq.(\ref{alpha_T}) will be to constrain standard Horndeski gravity models (and new models) against the multimessenger constraints on the propagation of GWs \cite{TheLIGOScientific:2017qsa} in the context of solving for the $G_{\text{Tele}}$ contribution. Other phenomenological effects may produce further restrictions involving the new contribution. Given the large number of new scalars in this new framework, it is crucial to determine whether these observational constraints can produce models that are consistent with current observations.

The impact of the modification parameters in Eq.(\ref{BDLS_GWPE}) is that the waveform will be altered both in amplitude and phase by the $\alpha_{M}$ and $\alpha_{T}$ parameters respectively. Expanding the waveform about its GR limit gives \cite{Ezquiaga:2018btd,PhysRevD.97.104037}
\begin{equation}
h_{\text{BDLS}}\sim h_{\rm GR}\;\underset{{\scriptscriptstyle \rm Amplitude}}{\underbrace{e^{-\frac{1}{2}\int\alpha_{M}\mathcal{H}d\eta}}}\;\underset{{\scriptscriptstyle \rm Phase}}{\underbrace{e^{ik\int\sqrt{\alpha_{T}+\frac{a^{2}\mu^{2}}{k^{2}}}d\eta}}}\,,
\end{equation}
where $\eta=\int dt/a$ denotes conformal time, $\mathcal{H}=a'/a$ is the conformal Hubble parameter, $\mu$ is an effective mass, and primes represent derivatives with respect to conformal time. A direct consequence of this modification to the GWPE is that the GW luminosity distance will also be effected \cite{Ezquiaga:2018btd,Belgacem:2019pkk}. BDLS theory generalized the standard gravity Horndeski theory by considering the TG analog of the same theory. However, the theory can be forced to produce no tensor excess speed, i.e. $\alpha_T=0$. In these cases, the luminosity distance for the GWs is related to their electromagnetic counterpart by \cite{Ezquiaga:2017ekz}
\begin{equation}
\frac{d_L^{g}(z)}{d_L^{EM}(z)} = \exp\left[\frac{1}{2}\int_0^z \frac{\alpha_M}{1+z'}\,dz'\right]\,,
\end{equation}
from which the damping of GWs against $z$ can be used to constrain the frictional term $\alpha_M$. This can be done using standard sirens which is one of the main aims of the next generation of gravitational wave detectors.

\section{Reviving Horndeski using Teleparallel gravity\label{Reviving_HG_Examples}}

From recent GW observations~\cite{TheLIGOScientific:2017qsa}, it was found that the speed of the gravitational waves is constrained to Eq.~(\ref{GW_constraint}). This equation effectively sets $\alpha_T \approx 0$ in a flat cosmological background. For the standard Horndeski case ($G_{\rm Tele}=0$), from Eq.~(\ref{alpha_T}), one can notice that in order to achieve this condition, one requires $G_4(\phi,X)=G_4(\phi)$ and $G_5(\phi,X)=\textrm{const.}$ (trivial). In greater detail, quartic and quintic Galileon models \cite{Nicolis:2008in,Deffayet:2009wt}, de-Sitter Horndeski \cite{Martin-Moruno:2015bda}, the Fab Four \cite{Charmousis:2011bf}, as well as purely kinetic coupled models~\cite{Gubitosi:2011sg} are severely constrained due to Eq.~(\ref{GW_constraint}). Indicatively, for example, the theory that reads 
\begin{align}\label{nonminimalcoupling}
    \mathcal{S} = & \int d^4x \sqrt{-g} \Big\{\frac{\mathring{R}}{\kappa ^2} - \left[ \epsilon\, g_{\mu\nu} + \eta\, \mathring{G}_{\mu\nu} \right]\phi^{;\mu}\phi^{;\nu} - 2 V(\phi) \Big\}+ \mathcal{S}_{\rm matter}\,,
\end{align}
where $\epsilon$ and $\eta$ are two coupling constants and $\mathring{G}_{\mu\nu}$ is the Einstein tensor, gives great phenomenology at different cosmological epochs because of the presence of the nonminimal kinetic coupling. It was very well studied in the literature~\cite{Sushkov:2009hk,Starobinsky:2016kua,Matsumoto:2017gnx,Sushkov:2012za,Amendola:1993uh,Saridakis:2010mf,Gubitosi:2011sg,Capozziello:2018gms} since it provides a realistic cosmological scenario emanated from this higher-order coupling. Namely, at early times it gives a quasi de Sitter behavior for the scale factor as an inflationary scenario; once inflation is over the Universe enters a matter-dominated era and later on, because of the dominance of the cosmological terms, it obtains a de Sitter behavior. The change between the epochs happens naturally without any fine-tuning potential. In greater detail, it is known that the universe in the model  \eqref{nonminimalcoupling}, depending on the coupling parameter, transits from one de Sitter solution to another and one can obtain ``a big bang, an expanding universe without a beginning, a cosmological turnaround, an eternally contractive universe, a big crunch, a big rip avoidance and a cosmological bounce'' \cite{Saridakis:2010mf}. Furthermore, in \cite{Matsumoto:2017gnx} dynamical analysis of \eqref{nonminimalcoupling} shows that there exist attractors representing three accelerated regimes of the Universe evolution, including de Sitter expansion and the little and big rip scenarios.

Another example is the so-called quartic Galileon model. Its action reads
\begin{equation}\label{quartic}
    \mathcal{S} = \int d^4x \sqrt{-g}\left[\frac{\mathring{R}}{\kappa ^2} + \sum _{i=1} ^4 \mathcal{L}_{i}\right] \,,
\end{equation}
where $\mathcal{L}_i$ are the known functions of the Horndeski theory~\cite{PhysRevD.100.064018,Capozziello:2018gms}. This model is very well studied as well; in~Ref.\cite{Gannouji:2010au} the authors found self-accelerating solutions, and they studied their stability, as well as spherically symmetric solutions. In~Ref.\cite{Li:2013tda}, they perform simulations showing that the Vainshtein mechanism suppresses very efficiently the spatial variations of the scalar field and in addition, the simulations fit very well both CMB and BAO data. In \cite{Bolis:2018kcq} they study the so-called parametrized post-Newtonian-Vainshteinian (PPNV) formalism of the quartic \eqref{quartic} and the quintic Galileon, that is an extension to the known PPN formalism and it is aimed to theories that need the Vainshtein mechanism to screen out the scalar field. Furthermore, in \cite{Elvang:2017mdq} the show that the model \eqref{quartic} can be supersymmetrized using the Galileon shift symmetry for the scalar and an ordinary shift symmetry for the fermionic sector. 

However, after the observation of GW170817, such nonminimal couplings Eq.~\eqref{nonminimalcoupling} and also models like Eq.~\eqref{quartic} were eliminated by the constraint in Eq.~\eqref{GW_constraint}, predicting a higher than the speed of light speed for the gravitational waves. 

In BDLS theory, when one assumes $G_{\rm Tele}\neq 0$, it is possible to find a theory which satisfies $\alpha_{T}=0$. To find a theory respecting that GWs must propagate at $c$, we need impose that $\alpha_T=0$. Then, for the BDLS theory, we impose that Eq.~\eqref{alpha_T} is equal to zero and then find out the corresponding functions $G_{\rm Tele},\, G_{\rm 4}$ and $G_{\rm 5}$ which ensure this condition. If one imposes this [$\alpha_T=0$ in~\eqref{alpha_T}], one finds that $G_5=G_5(\phi)$ and $G_{\rm Tele}=\tilde{G}_{\rm tele}(\phi ,X,T,T_{\rm vec},T_{\rm ax},I_2,J_1,J_3,J_6,J_8-4J_5,J_{10})$, which effectively gives that the BDLS Lagrangian satisfying the property that the propagation of the GW is equal to the speed of light is
\begin{widetext}
\begin{align}
  \mathcal{L}&=\tilde{G}_{\rm tele}(\phi ,X,T,T_{\rm vec},T_{\rm ax},I_2,J_1,J_3,J_6,J_8-4J_5,J_{10})+G_2(\phi,X)+G_3(\phi,X)\Box \phi\,,\nonumber\\
   &+G_4(\phi,X)\left(-T+B\right)+G_{4,X}\left[\left(\Box\phi\right)^2 - \phi_{;\mu\nu}\phi^{;\mu\nu}+4J_5\right]+G_5(\phi)\mathcal{G}_{\mu\nu}\phi^{;\mu\nu}-4J_5G_{5,\phi}\,.
   \label{revivedLagrangian}
\end{align}
\end{widetext}
This is the most important result of this paper since the above Lagrangian contains nontrivial coupling functions $G_4(\phi,X)$ and $G_5(\phi)$ that were previously ruled out for the standard Horndeski case. One can notice that the Lagrangians $\mathcal{L}_4$ and $\mathcal{L}_5$ are now corrected by a term proportional to $J_5$; otherwise $c_T$ will not be one (or $\alpha_T=0$). With these corrections, models that were eliminated in standard Horndeski will survive in this framework. Specifically, as we can see from the last four terms in Eq.~\eqref{revivedLagrangian}, theories with $G_5(\phi)$ and also $G_4(\phi,X)$ will give the correct speed for the gravitational waves. This correction of course includes the models in Eqs.~\eqref{nonminimalcoupling} and~\eqref{quartic}.

\section{Conclusions \label{Conclusion}}

In Ref.~\cite{PhysRevD.100.064018}, we introduced the Teleparallel analog of Horndeski, which relies upon the torsion tensor instead of the curvature tensor. This theory was built using the same conditions as in standard Horndeski, which are: (i) field equations must be at most second order in tetrad derivatives; (ii) the theory must be not parity violating. Due to the mathematical nature of the torsion tensor, it is possible to construct infinite scalars leading to second-order field equations, so that we also added an additional condition: (iii) we considered only contractions of the torsion tensor only up to quadratic terms. As another implicit condition, our theory is local Lorentz invariance. We saw that because of the structure of the torsion tensor, there appears a new function adding richer phenomenology to the theory. Hence, this theory can be written as Horndeski theory plus an additional term which comes from teleparallel scalars.

Horndeski theory is the most general scalar-tensor theory leading to second-order field equations. Most modified theories of gravity can be mapped onto its action. However, after the observation of GW170817, a significant part of the theory was eliminated because they predict discrepancies between the GW speed of propagation and the speed of light. In this work, we study the tensor perturbations of the tetrad in order to see whether there are models that revive the GW observation. To do this, we took a flat Friedmann-Lema\^{\i}tre-Robertson-Walker metric with its corresponding tetrad, and then we perform the tensorial cosmological perturbations. Since the tetrads have six extra degrees of freedom than the metric, we also checked that after considering local Lorentz transformations of our tetrad, the antisymmetric field equations vanish, as expected. Then, the tetrad used is compatible with a vanishing spin connection, and then, standard cosmological perturbations can be used for our BDLS theory. 

The most important result of this paper is given in Eq.~\eqref{alpha_T} where the excess tensor speed is displayed. By setting $G_{\rm Tele}=0$, we recover the standard result found in Horndeski theory. Now, for $G_{\rm Tele}\neq 0$, interestingly enough, because Eq.~\eqref{alpha_T} is modified, there appears a correction term both in $\mathcal{L}_4$ and in $\mathcal{L}_5$ (see Eq.~\eqref{revivedLagrangian}), and thus many significant models survive to the constraint $c_T=1$ (or $\alpha_T=0$). Explicitly, by setting this condition, one gets that the Lagrangian~\eqref{revivedLagrangian} is compatible with the current GW velocity. It may be noted that the terms $G_4$ and $G_5$ get corrections coming from the invariant $J_5$ which is related to contraction of derivatives of the scalar field and the tensorial part of the torsion tensor. We also pointed out some models that could be revived in the teleparallel analog of Horndeski. These theories have attracted some attention in the past in standard Horndeski, but they were almost discarded before. 

It would be interested to investigate large-scale structure constraints \cite{Creminelli:2018xsv,Creminelli:2019kjy} which may further refine physically viable choices of the BDLS Lagrangian. It would be also interesting to use standard sirens and also to use binary coalescence to put bounds in BDLS theory, as it was done in~\cite{DAgostino:2019hvh,Nunes:2019bjq}. Similarly as in~\cite{Soudi:2018dhv}, as another important study that could be done, is to analyze the polarization of gravitational waves in BDLS theory. All of these studies will be done later in the future.

\begin{acknowledgments}
The authors would like to acknowledge networking support by the COST Action GWverse CA16104. The authors would like to acknowledge networking support by the COST Action CA18108. This article is based upon work from CANTATA COST (European Cooperation in Science and Technology) action CA15117, EU Framework Programme Horizon 2020. S.B. is supported by Mobilitas Pluss N$^\circ$ MOBJD423 by the Estonian government.
\end{acknowledgments}

\appendix
\section{The BDLS Field Equations \label{BDLS_FEs}}
As in Ref.\cite{Bahamonde:2020cfv}, the field equations of the BDLS action in Eq.~(\ref{BDLS_Action}) can be determined by first taking a variation of this action with respect to the tetrad, which results in
\begin{eqnarray}
 \delta_{e}\mathcal{S}_{\rm BDLS}= e\mathcal{L}_{\rm Tele}e_a{}^{\mu}\delta e^a{}_\mu+e\delta_{e} \mathcal{L}_{\rm Tele}+e\sum_{i=2}^{5}\mathcal{L}_ie_a{}^{\mu}\delta e^a{}_\mu+e \delta_{e} \sum_{i=2}^{5}\mathcal{L}_{i}+2\kappa^2e\Theta_a{}^\mu\delta e^{a}{}_\mu=0\,,\label{deltaS}
\end{eqnarray}
where we have also included the standard minimally coupled matter Lagrangian, $\mathcal{L}_{\rm m}$, which produced the energy-momentum tensor through the definition
\begin{align}
\Theta_a{}^\mu=\frac{1}{e}\frac{\delta (e\mathcal{L}_{\rm m})}{\delta e^a{}_{\mu}}\,.
\end{align}
As one would expect, the variations of the standard Horndeski gravity contributions, $\delta_{e}\sum_{i=2}^{5} \mathcal{L}_i$, gives the standard Horndeski gravity field equations, whereas the the variation of $\delta_{e} \mathcal{L}_{\rm Tele}$ is related to the extra terms coming from TG. After doing several computations, one finds that the field equations can be written as
\begin{align}
&4(\partial_{\lambda}G_{\rm Tele,T})S_{a}\,^{\lambda\mu}+4e^{-1}\partial_{\lambda}(e S_{a}\,^{\lambda\mu})G_{\rm Tele,T}-4G_{\textrm{Tele},T}T^{\sigma}\,_{\lambda a}S_{\sigma}\,^{\mu\lambda}+4G_{\rm Tele,T}\omega^{b}{}_{a\nu}S_{b}{}^{\nu\mu}\nonumber\\
&-\phi_{;a}\Big[G_{\rm Tele,X}\phi^{;\mu}-G_{\rm Tele,I_2}v^\mu   -2G_{\rm Tele,J_1}a^{\mu}a_{j}\phi^{;j} +G_{\rm Tele,J_3}v_i t_{k}{}^{\mu i}\phi^{;k}-2G_{\rm Tele,J_5} t^{i\mu k}t_{ijk}\phi^{;j}\nonumber\\
&+2G_{\rm Tele,J_6}t_{ilk}t^\mu{}_M{}^i\phi^{;k}\phi^{;l}\phi^{;m}-2G_{\rm Tele,J_8}t_{ijk}t^{ij}{}^{\mu}\phi^{;k}-G_{\rm Tele,J_{10}}a^j \phi^{;i}\Big(\epsilon_{\mu jcd}t_i{}^{cd}+\epsilon_{ijcd}t^{\mu cd}\Big)  \Big]\nonumber\\
&+\frac{1}{3}\Big[M^i(\epsilon_{ib}{}^{cd}
e_c{}^{ \mu}  T^{b}{}_{ ad}
-\epsilon_{ib}{}^{cd}
e_d{}^{ \mu} \omega^b{}_{ ac})+e^{-1}\partial_\nu\Big(eM^i\epsilon_{ia}{}^{cd}   e_c{}^{ \nu}e_d{}^{ \mu}\Big)\Big]\nonumber\\
&-N^i(e_i{}^{ \mu} \omega^\rho{}_{a \rho}	- \omega^\mu{}_{ai} -	T^\mu{}_{ai}     	-	v_a	e_i{}^{ \mu})+e^{-1}\partial_\nu\Big(eN^i(e_a{}^{ \nu } e_i{}^{ \mu}  - e_a{}^{ \mu } e_i{}^{ \nu})\Big)\nonumber\\
&-O^{ijk}H_{ijka}{}^{\mu}+e^{-1}\partial_\nu\Big(eO^{ijk}L_{ijka}{}^{\mu\nu}\Big)- \mathcal{L}_{\rm Tele}e_a{}^{\mu}+2e_a{}^{\nu}g^{\mu\alpha}\sum_{i=2}^{5}\mathcal{G}^{(i)}{}_{\alpha\nu}=2\kappa^2\Theta_a{}^{\mu}\,,
\end{align}
where 
\begin{align}
    M^{i} &= 2G_{\rm Tele,T_{\rm ax}} a^i+2G_{\rm Tele,J_1}\phi^{;i}\phi^{;j}a_{j}+G_{\rm Tele,J_{10}}\epsilon_{a}{}^{i}{}_{cd}\phi^{;a}\phi^{;j}t_{j}{}^{cd}\,,\label{FEs_M}\\
    N^i &= 2G_{\rm Tele,T_{\rm vec}} v^i+G_{\rm Tele,I_2}\phi^{;i}+2G_{\rm Tele,J_2}\phi^{;i}\phi^{;j}v_{j}+G_{\rm Tele,J_3}\phi^{;k}\phi^{;j}t^{i}{}_ {kj}\,,\label{FEs_N}\\
    O^{ijk} &= G_{\rm Tele,J_3}\phi^{;j}\phi^{;k}v^{i}+2G_{\rm Tele,J_5}\phi^{;l}\phi^{;j}t^{i}{}_{l}{}^{k}+2G_{\rm Tele,J_6}\phi^{;j}\phi^{;k}\phi^{;l}\phi^{;m}t^{i}{}_{lm}+2G_{\rm Tele,J_8}\phi^{;l}\phi^{;k}t^{ij}{}_{l}\nonumber\\
    &+G_{\rm Tele,J_{10}}\epsilon_{ab}{}^{jk}\phi^{;a} \phi^{;b}\phi^{;i}\,,\label{FEs_O}
\end{align}
and
\begin{align}
    H_{ijka}{}^{\mu} &:= \frac{\partial t_{ijk}}{\partial e^a{}_{ \mu}} = \nonumber\\
    & \frac{1}{2}\Big[\omega_{iaj}e_k{}^{\mu}-\omega_{iak}e_j{}^{\mu}-T_{ija}e_k{}^{\mu}-T_{iak}e_j{}^\mu+\omega_{jai}e_k{}^{\mu}-\omega_{jak}e_i{}^{\mu}-T_{jia}e_k{}^{\mu}-T_{jak}e_i{}^\mu\Big]\nonumber\\
    & +\frac{1}{6}\Big[\eta_{ki}C_{ja}{}^{\mu}-\eta_{kj}C_{ia}{}^{\mu}-2\eta_{ij}C_{ka}{}^{\mu}+v_j D_{kia}{}^{\mu}-v_i D_{kja}{}^{\mu}-2v_k D_{ija}{}^{\mu}\Big]\,.\label{FEs_tensorial0}\\
    L_{ijka}{}^{\mu\nu} &:= \frac{\partial t_{ijk}}{\partial e^a{}_{ \mu,\nu}} = \nonumber\\
    & \frac{1}{2}\Big[\eta_{ai}(e_j{}^{\nu}e_k{}^{\mu}-e_j{}^{\mu}e_k{}^{\nu})+\eta_{aj}(e_i{}^{\nu}e_k{}^{\mu}-e_i{}^{\mu}e_k{}^{\nu})\Big]+\frac{1}{6}\Big[\eta_{ki}(e_a{}^{ \nu } e_j{}^{ \mu}  - e_a{}^{ \mu } e_j{}^{ \nu})\nonumber\\
    & -\eta_{kj}(e_a{}^{ \nu } e_i{}^{ \mu}  - e_a{}^{ \mu } e_i{}^{ \nu})-2\eta_{ij}(e_a{}^{ \nu } e_k{}^{ \mu}  - e_a{}^{ \mu } e_k{}^{ \nu})\Big]\,,\label{tensorial}\\
    C_{ia}{}^{\mu} &:= \frac{\partial v_i}{\partial e^a{}_{\mu}} = {e_i{}^{ \mu} \omega^\rho{}_{a \rho}	- \omega^\mu{}_{ai} -	T^\mu{}_{ai}     	-	v_a	e_i{}^{ \mu} }\,,\label{vector}\\
    D_{kia}{}^{\mu} &:= \frac{\partial \eta_{ki}}{\partial e^a{}_\mu}=\delta^b_i \eta_{ab}e^\mu{}_k+\delta^b_k \eta_{ab}e^\mu{}_i-\eta_{ai}e^\mu{}_k-\eta_{ka}e^\mu{}_i\,.
\end{align}
The terms $\mathcal{G}^{(i)}{}_{\alpha\nu}$  $\sum_{i=2}^{5}\mathcal{G}^{(i)}{}_{\mu\nu}$ were explicitly found in \cite{Capozziello:2018gms} (see Eqs.~(13a)-(13d) therein). This constitutes the tetrad field equations which produce the ten independent field equations as well as the six extra antisymmetric independent field equations due to the invariance under LLTs as discussed in Sec.~\ref{BDLS_Rev}.

On the other hand, the scalar field will also produce dynamical equations. By taking a variations of the action with respect to the scalar field results in a modified Klein Gordon equation given by
\begin{equation}
\mathring{\nabla}^\mu\Big(J_{\mu\rm -Tele}+\sum_{i=2}^{5}J^{i}_\mu\Big)=P_{\phi-\rm Tele}+\sum_{i=2}^{5}P_{\phi}^i\,,
\end{equation}
where $J_{\mu\rm -Tele}$ and $P_{\phi-\rm Tele}$ are defined as
\begin{eqnarray}
J_{\mu\rm -Tele}&=&-G_{\rm Tele,X}(\mathring{\nabla}_\mu\phi)+G_{\rm Tele,I_2} v_\mu+2G_{\rm Tele,J_1}a_\mu a^\nu\mathring{\nabla}_\nu \phi-G_{\rm Tele,J_3}v_\alpha t_{\mu}{}^{\nu\alpha}(\mathring{\nabla}_\nu\phi)\nonumber\\
&&-2G_{\rm Tele,J_5}t^{\beta\nu\alpha}t_{\beta\mu\alpha}(\mathring{\nabla}_\nu\phi)+2G_{\rm Tele,J_8}t^{\alpha\nu}{}_{\mu}t_{\alpha\nu}{}^{\beta}(\mathring{\nabla}_\beta\phi)-2G_{\rm Tele,J_6}t^{\nu\alpha\beta}t_{\mu}{}^{\sigma}{}_\nu(\mathring{\nabla}_\alpha\phi)(\mathring{\nabla}_\beta\phi)(\mathring{\nabla}_\sigma\phi)\,,\nonumber\\
&&-G_{\rm Tele,J_{10}} a^\nu (\mathring{\nabla}_\alpha \phi) (\epsilon^{\mu}{}_{\nu\rho\sigma}t^{\alpha\rho\sigma}+\epsilon^{\alpha}{}_{\nu\rho\sigma}t^{\mu\rho\sigma})\,,\label{Jtele2}\\
P_{\phi-\rm Tele}&=&G_{\rm Tele,\phi}\,.\label{PTele2}
\end{eqnarray}
Using the identity in Eq.(\ref{Ricc_Tor_iden}), it follows that $P_{\phi}^i$ is given by~\cite{Capozziello:2018gms}
\begin{subequations}\label{Pphi}
	\begin{align}
	P_{\phi}^{2} &= G_{2,\phi}\,,\\
	P_{\phi}^{3} &= \mathring{\nabla}_{\mu}G_{3,\phi}\mathring{\nabla}^{\mu} \phi \,,\\
	P_{\phi}^{4} &= G_{4,\phi}(-T+B) + G_{4,\phi X} \left[ (\mathring{\square} \phi)^2 - (\mathring{\nabla}_{\mu}\mathring{\nabla}_{\nu}\phi)^2\right]\,,\\
	P_{\phi}^{5} &= -\mathring{\nabla}_{\mu}G_{5,\phi} \mathring{G}^{\mu\nu}\mathring{\nabla}_{\nu}\phi - \frac{1}{6}G_{5,\phi X}\left[(\square \phi)^3 - 3 \square \phi (\mathring{\nabla}_{\mu}\mathring{\nabla}_{\nu} \phi)^2 + 2 (\mathring{\nabla}_{\mu}\mathring{\nabla}_{\nu}\phi)^3 \right]\,,
	\end{align}
\end{subequations}
whereas $J^{i}_\mu$ will be defined as
\begin{subequations}\label{Jis}
	\begin{align}
	J_{\mu}^{2} &= -\mathcal{L}_{2,X}\mathring{\nabla}_{\mu}\phi \,,\\
	J_{\mu}^{3} &= -\mathcal{L}_{3,X}\mathring{\nabla}_{\mu}\phi + G_{3,X} \mathring{\nabla}_{\mu}X + 2 G_{3,\phi} \mathring{\nabla}_{\mu} \phi \,,\\
	J_{\mu}^{4} &= - \mathcal{L}_{4,X}\mathring{\nabla}_{\mu} \phi +2 G_{4,X}\mathring{R}_{\mu\nu}\mathring{\nabla}^{\nu} \phi  - 2 G_{4,XX}\left(\mathring{\square} \phi \mathring{\nabla}_{\mu}X - \mathring{\nabla}^{\nu} X \mathring{\nabla}_{\mu}\mathring{\nabla}_{\nu} \phi \right) \nonumber \\
	&-2 G_{4,\phi X} (\mathring{\square} \phi \mathring{\nabla}_{\mu}\phi + \mathring{\nabla}_{\mu}X) \,,\\
	J_{\mu}^{5} &= -\mathcal{L}_{5,X}\mathring{\nabla}_{\mu}\phi - 2 G_{5,\phi}\mathring{G}_{\mu\nu} \mathring{\nabla}^{\nu} \phi \nonumber \\
	&-G_{5,X}\left[ \mathring{G}_{\mu\nu}\mathring{\nabla}^{\nu} X + \mathring{R}_{\mu\nu}\square \phi \mathring{\nabla}^{\nu} \phi - \mathring{R}_{\nu \lambda} \mathring{\nabla}^{\nu} \phi \mathring{\nabla}^{\lambda}\mathring{\nabla}_{\mu} \phi - \mathring{R}_{\alpha \mu \beta \nu}\mathring{\nabla}^{\nu}\phi \mathring{\nabla}^{\alpha} \mathring{\nabla}^{\beta}\phi\right]  \nonumber \\
	&+G_{5,XX} \Big\{ \frac{1}{2}\mathring{\nabla}_{\mu}X \left[(\mathring{\square} \phi)^2 - (\mathring{\nabla}_{\alpha}\mathring{\nabla}_{\beta}\phi)^2 \right]- \mathring{\nabla}_{\nu}X\left(\mathring{\square} \phi \mathring{\nabla}_{\mu}\mathring{\nabla}^{\nu}\phi - \mathring{\nabla}_{\alpha}\mathring{\nabla}_{\mu}\phi\mathring{\nabla}^{\alpha}\mathring{\nabla}^{\nu}\phi\right)\Big\}  \nonumber \\
	&+G_{5,\phi X} \Big\{ \frac{1}{2}\mathring{\nabla}_{\mu}\phi \left[(\mathring{\square} \phi)^2 - (\mathring{\nabla}_{\alpha}\mathring{\nabla}_{\beta}\phi)^2 \right] + \mathring{\square} \phi \mathring{\nabla}_{\mu}X -\mathring{\nabla}^{\nu}X \mathring{\nabla}_{\nu}\mathring{\nabla}_{\mu}\phi  \Big\}\,.
	\end{align}
\end{subequations}
To fully express all terms in the above equations in terms only depending on teleparallel quantities, one can use the following identities
\begin{align}
\mathring{R}^{\lambda}\,_{\mu\sigma\nu} &= \mathring{\nabla}_{\nu}K_{\sigma}{}^{\lambda}{}_{\mu} -
\mathring{\nabla}_{\sigma}K_{\nu}{}^{\lambda}{}_{\mu} +
K_{\sigma}{}^{\rho}{}_{\mu}K_{\nu}{}^{\lambda}{}_{\rho} -
K_{\sigma}{}^{\lambda}{}_{\rho}K_{\nu}{}^{\rho}{}_{\mu}\,,\\
\mathring{R}_{\mu\nu} &= \mathring{\nabla}_{\nu}K_{\lambda}{}^{\lambda}{}_{\mu} -
\mathring{\nabla}_{\lambda}K_{\nu}{}^{\lambda}{}_{\mu} +
K_{\lambda}{}^{\rho}{}_{\mu}K_{\nu}{}^{\lambda}{}_{\rho} -
K_{\lambda}{}^{\lambda}{}_{\rho}K_{\nu}{}^{\rho}{}_{\mu}\,,\\
\mathring{G}_{\mu\nu} &= e^{-1}e^{a}{}_{\mu}g_{\nu\rho}\partial_\sigma(e S_a{}^{\rho\sigma})-S_{b}{}^{\sigma}{}_{\nu}T^{b}{}_{\sigma\mu}+\frac{1}{4}T g_{\mu\nu}-e^{a}{}_\mu \omega ^{b}{}_{a\sigma}S_{b\nu}{}^{\sigma}\,.
\end{align}


\begin{thebibliography}{10}

\bibitem{Abbott:2016blz}
B.~P. Abbott et~al.
\newblock {Observation of Gravitational Waves from a Binary Black Hole Merger}.
\newblock {\em Phys. Rev. Lett.}, 116(6):061102, 2016.

\bibitem{TheLIGOScientific:2017qsa}
B.~P. Abbott et~al.
\newblock {GW170817: Observation of Gravitational Waves from a Binary Neutron
  Star Inspiral}.
\newblock {\em Phys. Rev. Lett.}, 119(16):161101, 2017.

\bibitem{Aldrovandi:2013wha}
Ruben Aldrovandi and José~Geraldo Pereira.
\newblock {\em {Teleparallel Gravity}}, volume 173.
\newblock Springer, Dordrecht, 2013.

\bibitem{Amendola:1993uh}
Luca Amendola.
\newblock {Cosmology with nonminimal derivative couplings}.
\newblock {\em Phys. Lett.}, B301:175--182, 1993.

\bibitem{Bahamonde:2017wwk}
Sebastian Bahamonde, Christian~G. Böhmer, and Martin Krššák.
\newblock {New classes of modified teleparallel gravity models}.
\newblock {\em Phys. Lett.}, B775:37--43, 2017.

\bibitem{Bahamonde:2015zma}
Sebastian Bahamonde, Christian~G. Bohmer, and Matthew Wright.
\newblock {Modified teleparallel theories of gravity}.
\newblock {\em Phys. Rev.}, D92(10):104042, 2015.

\bibitem{Bahamonde:2020cfv}
Sebastian Bahamonde, Konstantinos~F. Dialektopoulos, Manuel Hohmann, and
  Jackson Levi~Said.
\newblock {Post-Newtonian limit of Teleparallel Horndeski gravity}.
\newblock 2020.

\bibitem{PhysRevD.100.064018}
Sebastian Bahamonde, Konstantinos~F. Dialektopoulos, and Jackson~Levi Said.
\newblock Can horndeski theory be recast using teleparallel gravity?
\newblock {\em Phys. Rev. D}, 100:064018, Sep 2019.

\bibitem{Bardeen:1980kt}
James~M. Bardeen.
\newblock {Gauge Invariant Cosmological Perturbations}.
\newblock {\em Phys. Rev.}, D22:1882--1905, 1980.

\bibitem{Belgacem:2019pkk}
Enis Belgacem et~al.
\newblock {Testing modified gravity at cosmological distances with LISA
  standard sirens}.
\newblock {\em JCAP}, 1907(07):024, 2019.

\bibitem{Bellini:2014fua}
Emilio Bellini and Ignacy Sawicki.
\newblock {Maximal freedom at minimum cost: linear large-scale structure in
  general modifications of gravity}.
\newblock {\em JCAP}, 1407:050, 2014.

\bibitem{Bolis:2018kcq}
Nadia Bolis, Constantinos Skordis, Daniel~B. Thomas, and Tom Złośnik.
\newblock {Parametrized post-Newtonian-Vainshteinian formalism for the Galileon
  field}.
\newblock {\em Phys. Rev.}, D99(8):084009, 2019.

\bibitem{Capozziello:2018gms}
Salvatore Capozziello, Konstantinos~F. Dialektopoulos, and Sergey~V. Sushkov.
\newblock {Classification of the Horndeski cosmologies via Noether Symmetries}.
\newblock {\em Eur. Phys. J.}, C78(6):447, 2018.

\bibitem{Chan_2018}
Man~Ho Chan and Hon~Ka Hui.
\newblock Testing the cubic galileon gravity model by the milky way rotation
  curve and sparc data.
\newblock {\em The Astrophysical Journal}, 856(2):177, Apr 2018.

\bibitem{Charmousis:2011bf}
Christos Charmousis, Edmund~J. Copeland, Antonio Padilla, and Paul~M. Saffin.
\newblock {General second order scalar-tensor theory, self tuning, and the Fab
  Four}.
\newblock {\em Phys. Rev. Lett.}, 108:051101, 2012.

\bibitem{Clifton:2011jh}
Timothy Clifton, Pedro~G. Ferreira, Antonio Padilla, and Constantinos Skordis.
\newblock {Modified Gravity and Cosmology}.
\newblock {\em Phys. Rept.}, 513:1--189, 2012.

\bibitem{PhysRevLett.122.061301}
Edmund~J. Copeland, Michael Kopp, Antonio Padilla, Paul~M. Saffin, and
  Constantinos Skordis.
\newblock Dark energy after gw170817 revisited.
\newblock {\em Phys. Rev. Lett.}, 122:061301, Feb 2019.

\bibitem{Creminelli:2018xsv}
Paolo Creminelli, Matthew Lewandowski, Giovanni Tambalo, and Filippo Vernizzi.
\newblock {Gravitational Wave Decay into Dark Energy}.
\newblock {\em JCAP}, 1812(12):025, 2018.

\bibitem{Creminelli:2019kjy}
Paolo Creminelli, Giovanni Tambalo, Filippo Vernizzi, and Vicharit
  Yingcharoenrat.
\newblock {Dark-Energy Instabilities induced by Gravitational Waves}.
\newblock 2019.

\bibitem{Creminelli:2017sry}
Paolo Creminelli and Filippo Vernizzi.
\newblock {Dark Energy after GW170817 and GRB170817A}.
\newblock {\em Phys. Rev. Lett.}, 119(25):251302, 2017.

\bibitem{DAgostino:2019hvh}
Rocco D'Agostino and Rafael~C. Nunes.
\newblock {Probing observational bounds on scalar-tensor theories from standard
  sirens}.
\newblock {\em Phys.\ Rev.\ D}, 100(4):044041, 2019.

\bibitem{deRham:2016nuf}
Claudia de~Rham, J.~Tate Deskins, Andrew~J. Tolley, and Shuang-Yong Zhou.
\newblock {Graviton Mass Bounds}.
\newblock {\em Rev. Mod. Phys.}, 89(2):025004, 2017.

\bibitem{deRham:2018red}
Claudia de~Rham and Scott Melville.
\newblock {Gravitational Rainbows: LIGO and Dark Energy at its Cutoff}.
\newblock {\em Phys. Rev. Lett.}, 121(22):221101, 2018.

\bibitem{Deffayet:2009wt}
C.~Deffayet, Gilles Esposito-Farese, and A.~Vikman.
\newblock {Covariant Galileon}.
\newblock {\em Phys. Rev.}, D79:084003, 2009.

\bibitem{Elvang:2017mdq}
Henriette Elvang, Marios Hadjiantonis, Callum R.~T. Jones, and Shruti
  Paranjape.
\newblock {On the Supersymmetrization of Galileon Theories in Four Dimensions}.
\newblock {\em Phys. Lett.}, B781:656--663, 2018.

\bibitem{Ezquiaga:2017ekz}
Jose~María Ezquiaga and Miguel Zumalacarregui.
\newblock {Dark Energy After GW170817: Dead Ends and the Road Ahead}.
\newblock {\em Phys. Rev. Lett.}, 119(25):251304, 2017.

\bibitem{Ezquiaga:2018btd}
Jose~María Ezquiaga and Miguel Zumalacárregui.
\newblock {Dark Energy in light of Multi-Messenger Gravitational-Wave
  astronomy}.
\newblock {\em Front. Astron. Space Sci.}, 5:44, 2018.

\bibitem{Gannouji:2010au}
Radouane Gannouji and M.~Sami.
\newblock {Galileon gravity and its relevance to late time cosmic
  acceleration}.
\newblock {\em Phys. Rev.}, D82:024011, 2010.

\bibitem{Gleyzes:2013ooa}
Jerome Gleyzes, David Langlois, Federico Piazza, and Filippo Vernizzi.
\newblock {Essential Building Blocks of Dark Energy}.
\newblock {\em JCAP}, 1308:025, 2013.

\bibitem{Gleyzes:2014dya}
Jérôme Gleyzes, David Langlois, Federico Piazza, and Filippo Vernizzi.
\newblock {Healthy theories beyond Horndeski}.
\newblock {\em Phys. Rev. Lett.}, 114(21):211101, 2015.

\bibitem{Goldstein:2017mmi}
A.~Goldstein et~al.
\newblock {An Ordinary Short Gamma-Ray Burst with Extraordinary Implications:
  Fermi-GBM Detection of GRB 170817A}.
\newblock {\em Astrophys. J.}, 848(2):L14, 2017.

\bibitem{Gonzalez:2015sha}
P.~A. Gonzalez and Yerko Vasquez.
\newblock {Teleparallel Equivalent of Lovelock Gravity}.
\newblock {\em Phys. Rev.}, D92(12):124023, 2015.

\bibitem{Gubitosi:2011sg}
Giulia Gubitosi and Eric~V. Linder.
\newblock {Purely Kinetic Coupled Gravity}.
\newblock {\em Phys. Lett.}, B703:113--118, 2011.

\bibitem{PhysRevD.19.3524}
Kenji Hayashi and Takeshi Shirafuji.
\newblock New general relativity.
\newblock {\em Phys. Rev. D}, 19:3524--3553, Jun 1979.

\bibitem{Hohmann:2019nat}
Manuel Hohmann, Laur Jarv, Martin Krssak, and Christian Pfeifer.
\newblock {Modified teleparallel theories of gravity in symmetric spacetimes}.
\newblock {\em Phys. Rev.}, D100(8):084002, 2019.

\bibitem{Horndeski:1974wa}
Gregory~Walter Horndeski.
\newblock {Second-order scalar-tensor field equations in a four-dimensional
  space}.
\newblock {\em Int. J. Theor. Phys.}, 10:363--384, 1974.

\bibitem{Kobayashi:2019hrl}
Tsutomu Kobayashi.
\newblock {Horndeski theory and beyond: a review}.
\newblock {\em Rept. Prog. Phys.}, 82(8):086901, 2019.

\bibitem{Kobayashi:2011nu}
Tsutomu Kobayashi, Masahide Yamaguchi, and Jun'ichi Yokoyama.
\newblock {Generalized G-inflation: Inflation with the most general
  second-order field equations}.
\newblock {\em Prog. Theor. Phys.}, 126:511--529, 2011.

\bibitem{Koyama:2015vza}
Kazuya Koyama.
\newblock {Cosmological Tests of Modified Gravity}.
\newblock {\em Rept. Prog. Phys.}, 79(4):046902, 2016.

\bibitem{Krssak:2018ywd}
M.~Krssak, R.~J. Van Den~Hoogen, J.~G. Pereira, C.~G. Boehmer, and A.~A. Coley.
\newblock {Teleparallel Theories of Gravity: Illuminating a Fully Invariant
  Approach}.
\newblock 2018.

\bibitem{Krssak:2015oua}
Martin Kr\v{s}\v{s}\'{a}k and Emmanuel~N. Saridakis.
\newblock {The covariant formulation of f(T) gravity}.
\newblock {\em Class. Quant. Grav.}, 33(11):115009, 2016.

\bibitem{Li:2013tda}
Baojiu Li, Alexandre Barreira, Carlton~M. Baugh, Wojciech~A. Hellwing, Kazuya
  Koyama, Silvia Pascoli, and Gong-Bo Zhao.
\newblock {Simulating the quartic Galileon gravity model on adaptively refined
  meshes}.
\newblock {\em JCAP}, 1311:012, 2013.

\bibitem{Li:2010cg}
Baojiu Li, Thomas~P. Sotiriou, and John~D. Barrow.
\newblock {$f(T)$ gravity and local Lorentz invariance}.
\newblock {\em Phys. Rev.}, D83:064035, 2011.

\bibitem{Lovelock:1971yv}
D.~Lovelock.
\newblock {The Einstein tensor and its generalizations}.
\newblock {\em J. Math. Phys.}, 12:498--501, 1971.

\bibitem{Malik:2008im}
Karim~A. Malik and David Wands.
\newblock {Cosmological perturbations}.
\newblock {\em Phys. Rept.}, 475:1--51, 2009.

\bibitem{Martin-Moruno:2015bda}
Prado Martin-Moruno, Nelson~J. Nunes, and Francisco S.~N. Lobo.
\newblock {Horndeski theories self-tuning to a de Sitter vacuum}.
\newblock {\em Phys. Rev.}, D91(8):084029, 2015.

\bibitem{Matsumoto:2017gnx}
Jiro Matsumoto and Sergey~V. Sushkov.
\newblock {General dynamical properties of cosmological models with nonminimal
  kinetic coupling}.
\newblock {\em JCAP}, 1801(01):040, 2018.

\bibitem{misner1973gravitation}
C.W. Misner, K.S. Thorne, and J.A. Wheeler.
\newblock {\em Gravitation}.
\newblock Number pt. 3 in Gravitation. W. H. Freeman, 1973.

\bibitem{Mukhanov:1990me}
Viatcheslav~F. Mukhanov, H.~A. Feldman, and Robert~H. Brandenberger.
\newblock {Theory of cosmological perturbations. Part 1. Classical
  perturbations. Part 2. Quantum theory of perturbations. Part 3. Extensions}.
\newblock {\em Phys. Rept.}, 215:203--333, 1992.

\bibitem{Nicolis:2008in}
Alberto Nicolis, Riccardo Rattazzi, and Enrico Trincherini.
\newblock {The Galileon as a local modification of gravity}.
\newblock {\em Phys. Rev.}, D79:064036, 2009.

\bibitem{PhysRevD.97.104037}
Atsushi Nishizawa.
\newblock Generalized framework for testing gravity with gravitational-wave
  propagation. i. formulation.
\newblock {\em Phys. Rev. D}, 97:104037, May 2018.

\bibitem{Nunes:2019bjq}
Rafael~C. Nunes, Marcio E.~S. Alves, and Jose C.~N. de~Araujo.
\newblock {Forecast constraints on $f(T)$ gravity with gravitational waves from
  compact binary coalescences}.
\newblock 2019.

\bibitem{Nunes:2018zot}
Rafael~C. Nunes, Marcio E.~S. Alves, and Jose C.~N. de~Araujo.
\newblock {Primordial gravitational waves in Horndeski gravity}.
\newblock {\em Phys. Rev.}, D99(8):084022, 2019.

\bibitem{peacock1999cosmological}
J.A. Peacock.
\newblock {\em Cosmological Physics}.
\newblock Cambridge Astrophysics. Cambridge University Press, 1999.

\bibitem{Perenon:2015sla}
Louis Perenon, Federico Piazza, Christian Marinoni, and Lam Hui.
\newblock {Phenomenology of dark energy: general features of large-scale
  perturbations}.
\newblock {\em JCAP}, 1511(11):029, 2015.

\bibitem{Riazuelo:2000fc}
Alain Riazuelo and Jean-Philippe Uzan.
\newblock {Quintessence and gravitational waves}.
\newblock {\em Phys. Rev.}, D62:083506, 2000.

\bibitem{Rinaldi:2016oqp}
Massimiliano Rinaldi.
\newblock {Mimicking dark matter in Horndeski gravity}.
\newblock {\em Phys. Dark Univ.}, 16:14--21, 2017.

\bibitem{Sakstein:2017xjx}
Jeremy Sakstein and Bhuvnesh Jain.
\newblock {Implications of the Neutron Star Merger GW170817 for Cosmological
  Scalar-Tensor Theories}.
\newblock {\em Phys. Rev. Lett.}, 119(25):251303, 2017.

\bibitem{Saltas:2014dha}
Ippocratis~D. Saltas, Ignacy Sawicki, Luca Amendola, and Martin Kunz.
\newblock {Anisotropic Stress as a Signature of Nonstandard Propagation of
  Gravitational Waves}.
\newblock {\em Phys. Rev. Lett.}, 113(19):191101, 2014.

\bibitem{Saridakis:2010mf}
Emmanuel~N. Saridakis and Sergey~V. Sushkov.
\newblock {Quintessence and phantom cosmology with non-minimal derivative
  coupling}.
\newblock {\em Phys. Rev.}, D81:083510, 2010.

\bibitem{Soudi:2018dhv}
Ismail Soudi, Gabriel Farrugia, Viktor Gakis, Jackson Levi~Said, and
  Emmanuel~N. Saridakis.
\newblock {Polarization of gravitational waves in symmetric teleparallel
  theories of gravity and their modifications}.
\newblock {\em Phys.\ Rev.\ D}, 100(4):044008, 2019.

\bibitem{SpurioMancini:2019rxy}
A.~Spurio~Mancini, F.~Köhlinger, B.~Joachimi, V.~Pettorino, B.~M. Schäfer,
  R.~Reischke, Edo van Uitert, S.~Brieden, M.~Archidiacono, and J.~Lesgourgues.
\newblock {KiDS+GAMA: Constraints on Horndeski gravity from combined
  large-scale structure probes}.
\newblock 2019.

\bibitem{Starobinsky:2016kua}
Alexei~A. Starobinsky, Sergey~V. Sushkov, and Mikhail~S. Volkov.
\newblock {The screening Horndeski cosmologies}.
\newblock {\em JCAP}, 1606(06):007, 2016.

\bibitem{Sushkov:2009hk}
Sergey~V. Sushkov.
\newblock {Exact cosmological solutions with nonminimal derivative coupling}.
\newblock {\em Phys. Rev.}, D80:103505, 2009.

\bibitem{Sushkov:2012za}
Sergey~V. Sushkov.
\newblock {Realistic cosmological scenario with non-minimal kinetic coupling}.
\newblock {\em Phys. Rev.}, D85:123520, 2012.

\bibitem{Tamanini:2012hg}
Nicola Tamanini and Christian~G. Bohmer.
\newblock {Good and bad tetrads in f(T) gravity}.
\newblock {\em Phys. Rev.}, D86:044009, 2012.

\end{thebibliography}
\end{document}